\documentclass[pdflatex]{article}
\usepackage{subeqnarray}

\usepackage{graphicx}
\usepackage{epstopdf}
\usepackage{fullpage}
\usepackage{setspace}
\usepackage{charter}

\usepackage{hyperref} 
\hypersetup{colorlinks=true,citecolor=blue, linkcolor=blue}

\usepackage[authoryear,round]{natbib}

\begin{document}

\title{A Reanalysis of North Pacific Sea Surface Temperatures Using State-Space
Techniques:  The PDO In A New Light
\author{ 
Roy Mendelssohn\thanks{Email: {\tt roy.mendelssohn@noaa.gov}}\\
Environmental Research Division\\
NOAA/NMFS/SWFSC\\
Pacific Grove, CA, 93950 USA\\
\vspace{7pt}
\and
Cindy Bessey\thanks{Email: {\tt cbessey@fiu.edu}}\\
College of Arts and Sciences\\
Florida International University\\
Miami, FL, 33139 USA}
\date{}
}

\maketitle

\begin{abstract}
North Pacific sea surface temperatures (SST), as used in
estimating the PDO, are reanalyzed using state-space
decomposition and subspace identification techniques.  The reanalysis presents a
very different picture of SST in this region. The first common
trend reflects a global warming signal.  The second common trend modifies this
for areas that underwent a sharper warming  (cooling) starting in the early
1970's.  This trend is also related to dynamics in the tropics and in Arctic sea
ice extent. The third common trend is a superposition of changes in
pressure centers on the long-term global warming signal.  The fourth common
trend is the trend that is contained in the original PDO series if analyzed by
state-space techniques,  and is identical to the trend in the North Pacific
High.  The first two common stochastic cycles capture the original PDO and
so-called "Victoria mode", showing that these series are dominated by stationary
behavior..
\end{abstract}
\section{Introduction}
\label{intro}
 Since its introduction, the Pacific Decadal Oscillation
\citep[PDO][]{Mantua:1997dp} has been credited
for most of the decadal-scale variability found in the North Pacific region, and
is a widely used index, often used to portray "regime shifts". The PDO was
derived from an Empirical Orthogonal Function (EOF) analysis of sea surface
temperature (SST) anomalies in the North Pacific (NPAC). The PDO is the first
EOF of the anomalies after removing the monthly global mean from each location,
and it explains 25\% of the variance of the resulting series in the region. This
latter fact is important, as the loadings in many areas are quite small, and
these areas will have a non-significant correlation with the PDO,  implying that
these regions will vary with the PDO on the order of random noise.  Moreover,
even in areas with relatively higher correlations with the PDO, there will be
many periods where the dynamics of these regions differ from that portrayed by
the PDO.

Further,  \cite{ISI:000171261600018}  and \cite{ISI:000188458400001} argue that the
variability in the PDO does not differ from that of an autoregressive series. 
As autoregressive series are stationary with 
a constant mean and a constant covariance structure, if the PDO is an
autoregressive series then at least in a statistical sense the dynamics of SST
in the North Pacific as represented by the PDO are not changing. A standard
"cartoon" of the North Pacific is a "white-noise" atmosphere driving a
"red-noise" ocean  and Pierce (2001) shows that such a model does indeed
produce the characteristic spatial patterns of the warm/cold periods associated
with the PDO.

To further test if the PDO is simply an autoregressive series,  we estimated a series of state-space decomposition models for the
PDO.  The state-space decomposition allows for a variety of different models to
be fit to the data, including models with and without a trend term, with and
without various stochastic cycles or AR terms, and then the "best" model from
these alternatives can be chosen using model selection criteria such as AIC or
BIC. The best fit model included a non-stationary trend, a nearly deterministic
seasonal term, and a stochastic cycle term.  The stochastic cycle contained the
most variance, but the trend is real and significant implying that the PDO when
analyzed properly is more than an autoregressive series \citep[see][]{Mendelssohn:2010:JSSOBK:v41i02, Schwing_Mendelssohn_Bograd_Overland_Wang_Ito_2010}.

\begin{figure}
\noindent  \includegraphics[width=6in]{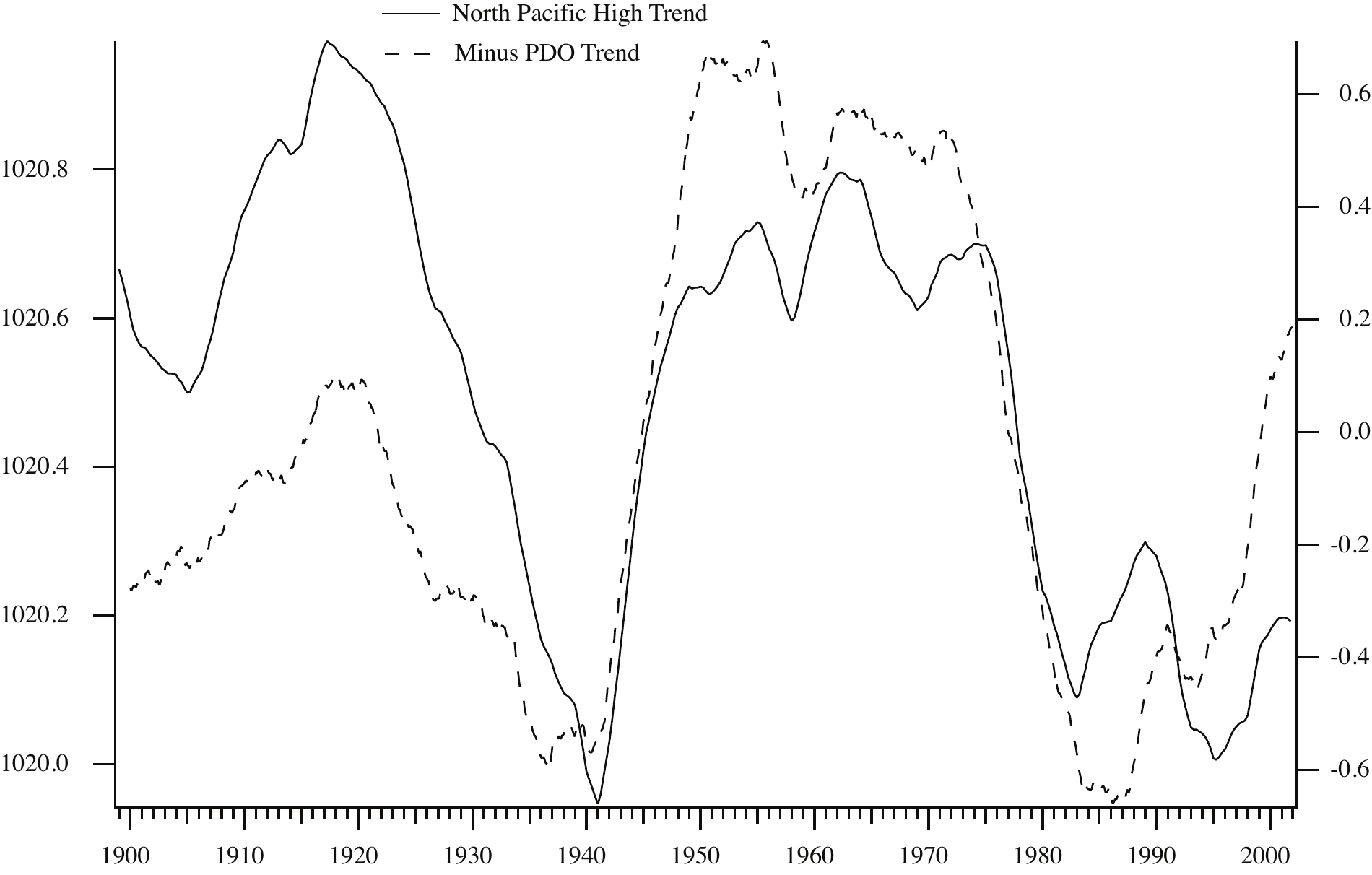}
 \caption{Comparison of (minus) the PDO trend (dashed line) with the estimated
trend for the North Pacific High (solid line).}
 \label{fig:pdo_trend}
\end{figure}

However, the trend term (Figure~\ref{fig:pdo_trend}) from this analysis is
almost identical to that from a state-space decomposition of  sea level pressure
(SLP) in the region of the North Pacific High  (NPH). Physically, it would seem
unlikely that the entire extra-tropical North Pacific SST is so heavily
influenced by the NPH, and therefore the PDO must be reflecting only part of the
variability in the region.

To properly estimate the dominant trends  in the extra-tropical North Pacific
SST series, and to separate these from the dominant stationary modes, requires a
reanalysis of these data. 
In this paper, we estimate "common trends" and "common cycles" in the NPAC  SST
data using the combination of state-space decompositions and subspace
identification techniques used in our previous papers  \citep{Mendelssohn:2002jl, Mendelssohn:2003um, Mendelssohn:2004wq,  ISI:000228479700003, Palacios:2004km} . Our results show that the trend in the original PDO discussed above
is identical to the fourth common trend in our analysis, and not surprisingly
the west coast of North America and the West Wind Drift region, 
regions that are influenced by the NPH, have the strongest weightings and are
separated out in the analysis. The first common trend appears to be a "global
warming" signal, and we present evidence that it is indeed a global, rather than
just a North Pacific signal. 

The second common trend modifies the first common trend in areas that underwent
a sharper warming  (cooling) starting in the early 1970's. The second common
trend is also similar to the long-term trend in El Ni\~{n}o related indices
identified previously in  \cite{ISI:000228479700003}  and to the trend in
sea ice extent in the North Pacific. 
The first two common stochastic cycles capture what is usually thought of as the
PDO and the "Victoria" mode, explaining the strong autoregressive behavior in
these indices. Finally, the higher-order common stochastic cycles suggest that
large regions oscillate in contrasting patterns through time.

As a methodological note, it is often asked why bother with the more complicated
state-space and subspace identification techniques instead of the usual EOF
analysis.  As the PDO is the first EOF of the SST series, our analysis provides
a clear example that the two techniques produce different results. 
State-space techniques also are more powerful, as it is possible with
state-space models to separate out stationary from non-stationary behavior, to
examine common behavior at different time-scales, and as there is a likelihood
function,  to compare models and to examine residuals for adequacy of the model
assumptions. 

\section{Data and Methods}
\label{sec:DM}
We tried to follow as closely as possible the data processing procedures
described by \cite{Mantua:1997dp}  in their analysis of the mid-latitude
North Pacific Basin SST.  We obtained monthly mean SST data for the years 1900
to 1993 from the U.K. Meteorological Office Global Sea Ice Coverage and SST
Dataset (HadISST 1.1), which is provided by the Hadley Centre. 
We also obtained the monthly mean Optimally Interpolated SST \citep[OISST.v2][]{Reynolds2002} data for years 1994 to present from the NOAA-CIRES
Climate Diagnostics Center (\url{http://www.cdc.noaa.gov}).  Both of these datasets
are on a 1$^{\circ}$ latitude by 1$^{\circ}$ longitude grid.  These datasets
were then combined and averaged into 5$^{\circ}$ boxes for our area of interest
; 20$^{\circ}$N - 65$^{\circ}$N and 110$^{\circ}$E Ð 250$^{\circ}$E. We refer to
boxes based on the latitude and longitude of the southwest corner, a convention
used in the COADS dataset. 

An index of the extent of sea ice coverage in the North Pacific was kindly
provided to us by William Chapman, and extends the index reported in 
\cite{Chapman1992}.  This index is quarterly for the years 1900-2005.

For each of the time series a state-space decomposition was calculated, and then
"common trends" were calculated from the estimated trends and "common cycles"
from the estimated stochastic cyclic terms. A state-space model decomposition
\citep{Harvey:1989bx, Durbin:2001wl}, which we have applied 
previously to examine long-term changes in the mean and seasonal components in
many other climate time series \citep[][and references therein]{Mendelssohn:2003um, Mendelssohn:2004wq} , allows for a series to be decomposed into a variety of
different independent components,  such as a nonparametric or fixed mean
(trend), autoregressive (AR) component, stochastic cycle, stochastic seasonal,
etc.:

\begin{equation}
      y(t) = T(t) + S(t) + I(t) + e(t), \qquad t=1,\tau, 
\end{equation} 

where, at time $t$, $T(t)$ is the unobserved time-dependent mean-level
(nonparametric trend), $S(t)$ is the seasonal component (zero-mean,
nonstationary and nondeterministic),  $I(t)$ is the irregular term (containing
any stationary autocorrelated or cyclic part of the data), and $e(t)$ is the
stationary uncorrelated component, which can be viewed here as "observation"
error. This model must be more fully specified to be meaningful, which is
accomplished by using piecewise continuous smoothing splines to estimate the
unobserved components.  \cite{Kita:Gers:smoo:1996},  \cite{Harvey:1989bx} 
and \cite{Durbin:2001wl} describe in detail how to parameterize the
model, and solve using a combination of Kalman filtering and smoothing, and
maximum likelihood estimation.

More specifically, the trend term can be viewed as a
unknown function of time, and parameterized as 
\begin{equation}
   \nabla^k T(t) \sim N(0,\sigma_T^2). 
   \label{trend_smooth} 
\end{equation} 
The seasonal component is usually constrained in one of two fashions. In the first, the
running sum of the seasonal component is constrained (assuming $s$ periods in
a season) 
\begin{equation} 
   \sum_{i=0}^{s-1} S(t-i) \sim N(0,\sigma_S^2), \qquad t=1,T, 
   \label{season1_smooth} 
\end{equation} 
while in the second the seasonal differences are constrained 
\begin{equation} 
   S(t) - S(t-s) \sim N(0,\sigma_S^2), \qquad t=1,T. 
   \label{season2_smooth} 
\end{equation} 

The state-space specification
of a stochastic cycle \citep{Durbin:2001wl}  is:
\begin{equation}
\left( 
\begin{array}{c}
   \psi_{t}  \\
   \psi_{t}^{*} 
   \end{array}\right) =
   \rho \left( 
\begin{array}{rr}
   \cos \lambda_{c} &  \sin \lambda_{c} \\
    -\sin \lambda_{c} &  \cos \lambda_{c} 
   \end{array}\right)
   \left( 
\begin{array}{c}
   \psi_{t-1}  \\
   \psi_{t-1}^{*} 
   \end{array}\right) +
   \left( 
\begin{array}{c}
   \kappa_{t}  \\
   \kappa_{t}^{*} 
   \end{array}\right), \qquad t=1,\ldots,T,
\end{equation}
where $\psi_{t}$ and $\psi_{t}^{*}$ are the states, $\lambda_{c}$ is the
frequency, in radians, in the range $0 <
\lambda_{c} \leq \pi $, $\kappa_{t}$ and $\kappa_{t}^{*}$ are two mutually
uncorrelated white noise disturbances with zero means and common variance
$\sigma_{\kappa}^{2}$, and $\rho $ is a damping factor. The damping factor
$\rho$ in (1) accounts for the time over which a higher amplitude event
(consider this to be a "shock" to the series) in the stochastic cycle  will
contribute to subsequent cycles. A stochastic cycle has changing amplitude and
phase, and becomes a first order autoregression if  $\lambda_{c}$ is $0$ or
$\pi$. 
Moreover, it can be shown that as $\rho \rightarrow 1$, then
$\sigma_{\kappa}^{2}
\rightarrow 0$ and the stochastic cycle reduces to the stationary deterministic
cycle:
\begin{equation}
\psi_{t}= \psi_{0}\cos\lambda_{c}t + \psi_{0}^{*}\sin\lambda_{c}t, \qquad
t=1,\ldots,T.
\end{equation}

The state-space decomposition belongs to the more general class of linear gaussian (state-space) models that  are amenable to solution using the Kalman filter/smoother as:

\begin{subeqnarray}
   y(t) &=& A(t)x(t) + v(t) \slabel{obs_eqn}\\ 
   x(t) &=& \Phi x(t-1) + w(t) \slabel{state_eqn}
\end{subeqnarray}
where the {\em observation equation} (Eq. \ref{obs_eqn}) has $y(t)$ a $q \times 1$-vector of
the observed data (in this case $q=1$), $A(t)$ is a $q \times p$ matrix which
relates the data to the unobserved components $x(t)$, which is a vector of
dimension $p \times 1$, and $v(t)$ is a $q \times 1$-vector of independent,
identically distributed gaussian random variables with $Ev(t) = 0$ and noise
covariance matrix
\begin{equation}
   R = E(v(t)v(t)^\prime).
\end{equation}

The evolution of the unobserved components or states $x(t)$ is governed by
the initial value $x(0)$ and the {\em state equation} (Eq. \ref{state_eqn}).
The matrix $\Phi$ is a $p \times p$ {\em transition matrix} and the $p \times
1$-vector $w(t)$ is another independent, identically distributed gaussian
random variable with $E(w(t)) = 0$ and
\begin{equation}
   Q = E(w(t)w(t)^\prime).
\end{equation}
The specification of the model is completed by assuming that $x(0)$ is also gaussian
 with $E(x(0)) = \mu$ and
\begin{equation}
   \Sigma = E(x(0) - \mu)(x(0) - \mu)^\prime.
\end{equation}
See \cite[][Chapter 6]{Shumway:2006ap} for further details on the state-space model. and how the unknown parameters can be estimated using the EM algorithm.

For multivariate series, "common trends", "common cycles", and "common seasonals" can be estimated by the appropriate specification of the transition matrix $\Phi$ and the observation matrix $A$  \citep[see for example][,Chapter 8]{Harvey:1989bx}. However, the computations are impractical for very large problems such as those of this paper.  Instead the unobserved components are calculated using subspace identification methods \citep{chiuso2004-1, Hannan:1988ea,  Larimore:1996qt, Larimore:2000wm, Van-Overschee:1996gc}. Subspace identification methods are based on Akaike's   canonical correlation method for system identification \citep{Akai:mark:1975},  and proceed by calculating  the singular value decomposition of the system Hankel matrix, and then the system matrices are estimated using what are essentially multivariate least-squares procedures.  \cite{Smith2000}  and \cite{Ninness2000} compare the properties of the two methods for estimating linear stochastic systems and how they relate.

Subspace identification methods are usually calculated on the raw data, but as \cite{Aoki:1988xw} has pointed out higher frequency modes tend to have more variance, and therefore can obscure lower frequency modes.  For this reason, "common trends" were calculated by removing from each series the estimated stochastic cycle and (possibly non-stationary) seasonal terms, with similar calculations being done for the "common cycles".

The spatial maps shown are the dynamic "factor loadings" from the observation
matrix $A$ in the Kalman filter representation of 
the common trends or common cycles, except for the map of the first common
trend, which shows correlations instead of factor loadings 
because the first common trend contains a scale effect in the factor loadings -
i.e. the series is not zero mean 

For our models, we estimated a trend, a stochastic seasonal and a stochastic
cyclic term \citep[see][]{ISI:000228479700003}. 
For the original PDO index we also estimate a Markov State Switching Model
(MSSM), which is often called a "Hidden Markov Model" or HMM - see for example
\cite{Hamilton_1989} or \cite{Shumway:1991sl}. An MSSM extends the
state-space model by assuming that there is a hidden two-regime process that
switches regimes following a Markov chain, and that the parameters of the
state-space model differ depending on which regime the process is in.  In
particular, we estimate a Markov-Switching Autoregressive Model (MSAR) that
assumes that there is "High" and "Low" state each of which follows an AR(2)
model, but the mean and parameters of the AR(2) model differ.  The model is
constrained such that the mean of the "High" state can be no lower than that of
the "Low" state.

The analyses were performed using the Finmetrics package for S-Plus \citep{zivot2006modeling}.  The state-space portion of this package is based on the
SsfPack software developed by \cite{RePEc:ect:emjrnl:v:2:y:1999:i:1:p:107-160}.

\section{Results}
\label{sec:R}

\subsection{Two Stationary Regimes?}
\label{sec:R.regimes}

The main goal of this paper is to use state-space models 
to re-examine the extent to which the  PDO captures the dominant long-term and
mid-term variability in SST in the North Pacific, but a related issue  is the
interpretation of the PDO as two "stationary regimes" with the zero-crossing of
the index represents the switching of the surface temperature regime in the
north Pacific. The "two-regime" interpretation implies either  that the
zero-crossing of the PDO has a physical meaning such that when the PDO crosses
zero the rest of the ecosystem shifts, or else that embedded in the behavior of
the series are two regimes with differing dynamics.  Fortunately, the latter
possibility can be modeled explicitly.

To examine this question further, the PDO is modeled using a Markov Switching
Autoregressive Model.  The model assumes that there are two regimes  (a "warm"
and "cool" regime),  and an unobserved underlying process switches between the
two regimes with a given probability following a Markov Chain model.  Each
regime is assumed to follow its own stationary AR(2) model, and the mean of the
"warm" regime is constrained to be higher than the mean of the "cool" regime.
The output includes parameter estimates for each of the AR(2) models, smoothed
estimates of the probability of being in either regime,  and smoothed estimates
of the state of the system.

\begin{figure}
\noindent  \includegraphics[width=6in]{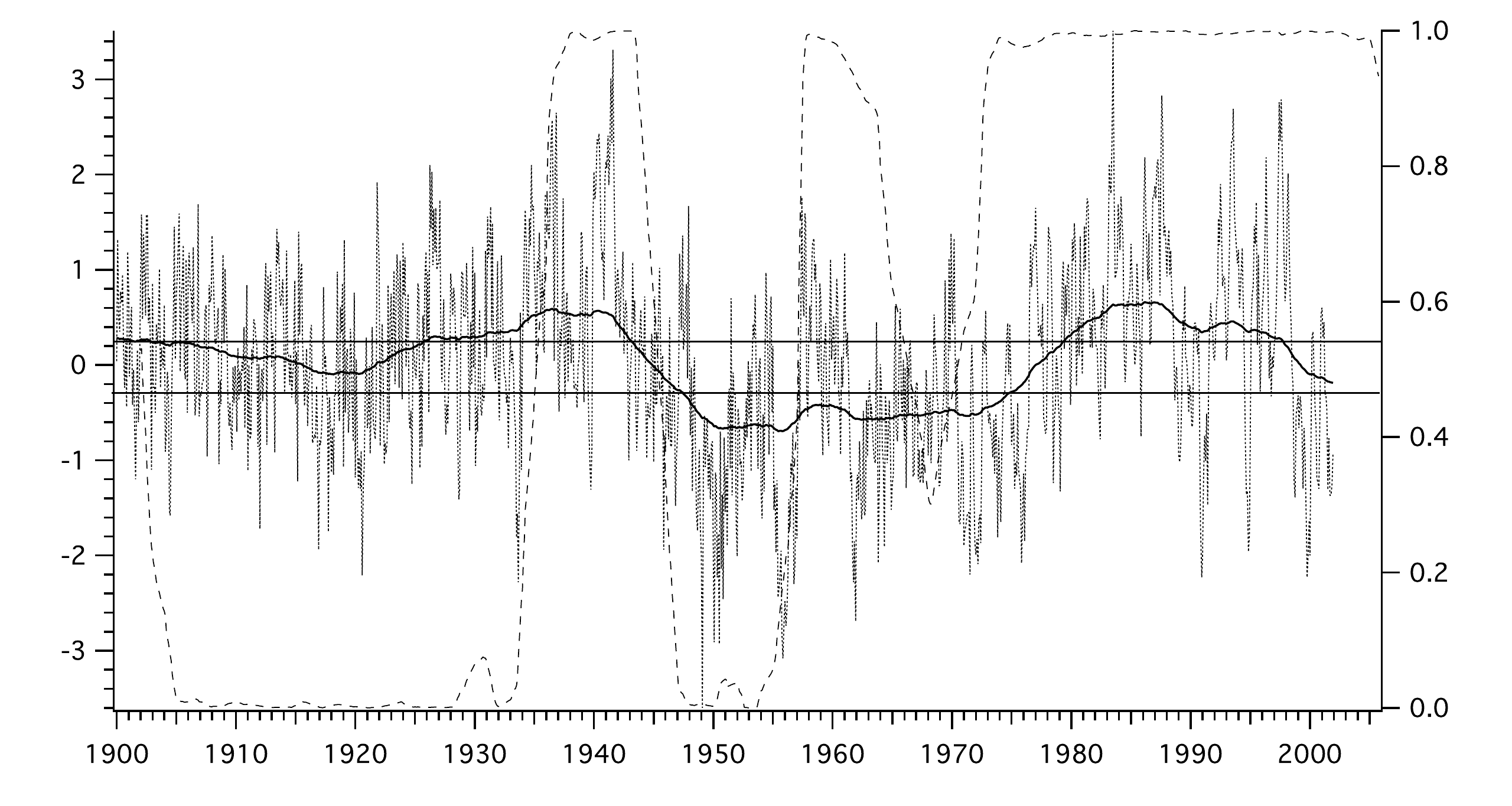}
 \caption{Results from a Markov-Switching Autoregressive Model of the PDO. The
PDO (dotted line) is compared with the estimated trend from a state-space
decomposition (solid line) and the smooth estimated probability of being in the
warm regime (dashed line). The estimated mean of the cool regime is not
significantly different from zero.}
 \label{fig:mssm}
\end{figure}

The output (Figure~\ref{fig:mssm}) suggests that if there were two regimes, then
the Pacific was in a warm regime only from the early 1930's to the late 1940's,
from 1956 to the early 1960's, and from 1970 onward. However,  there are reasons
to prefer the state-space decomposition to the switching model.  First the
state-space decomposition is more parsimonious, by the AIC criterion is a better
model and visually it explains the data better 
(particularly during periods such as 1984-1992 and after 1998). The means of the
two regimes barely differ (0.028, -0.037) and the mean of the cool regime is not
significantly different than zero, This suggests that a model with a single
mean, such as in the state-space decomposition, is to be preferred (though there
are some differences in the AR(2) model estimated for each regime, a possible
asymmetry which has been noted 
before), which is reinforced by the fact that the smoothed state-space residuals
do not show a lack of fit based on a variety of tests.

\subsection{Common Trends}
\label{sec:R.trends}

\begin{figure}
\noindent  \includegraphics[width=6in]{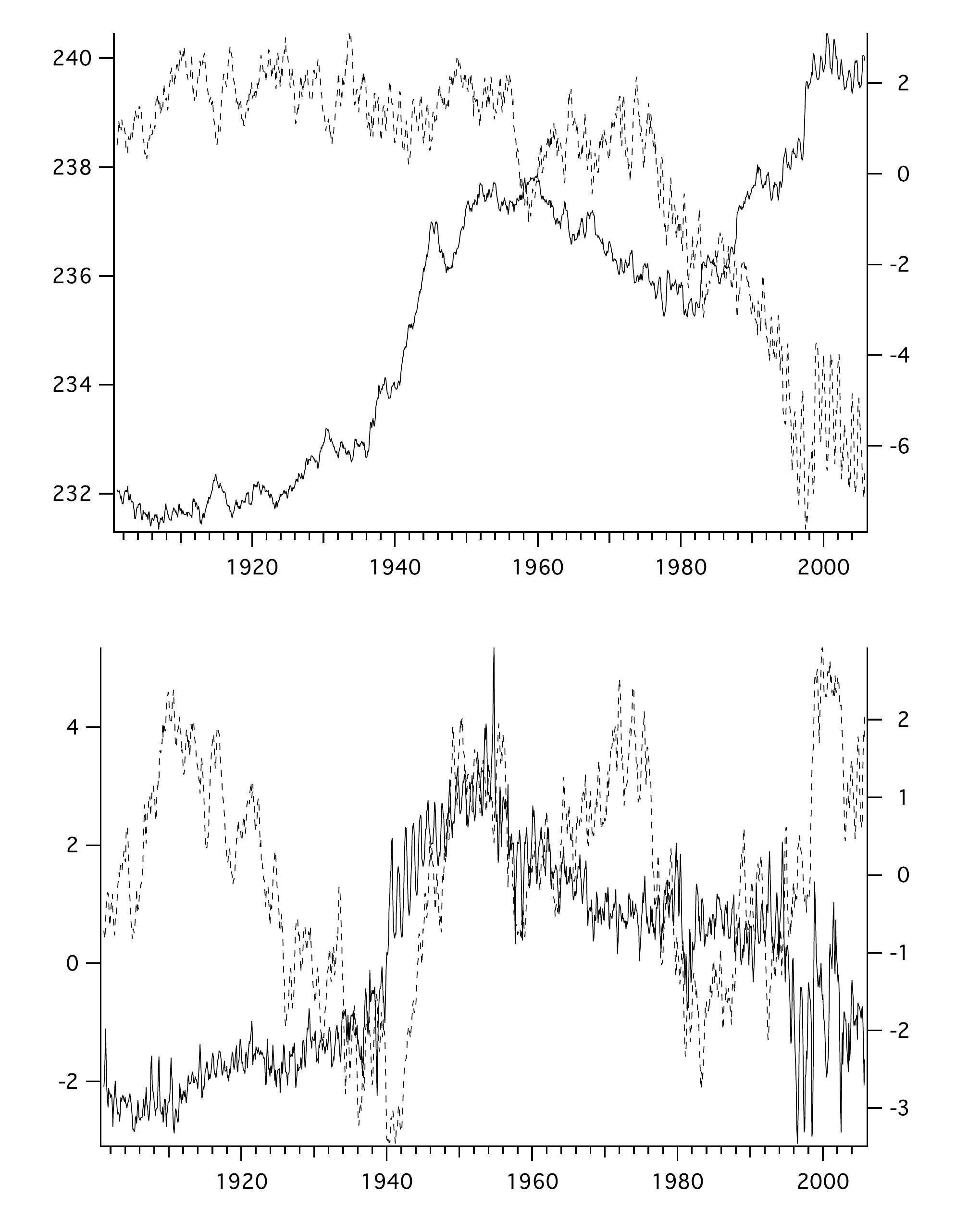}
 \caption{a). Common trend 1(solid line) and common trend 2 (dashed line). 
b). Common trend 3 (solid line) and common trend 4 (dashed line).}
 \label{fig:common_trend1234}
\end{figure}

\begin{figure}
\noindent  \includegraphics[width=6in]{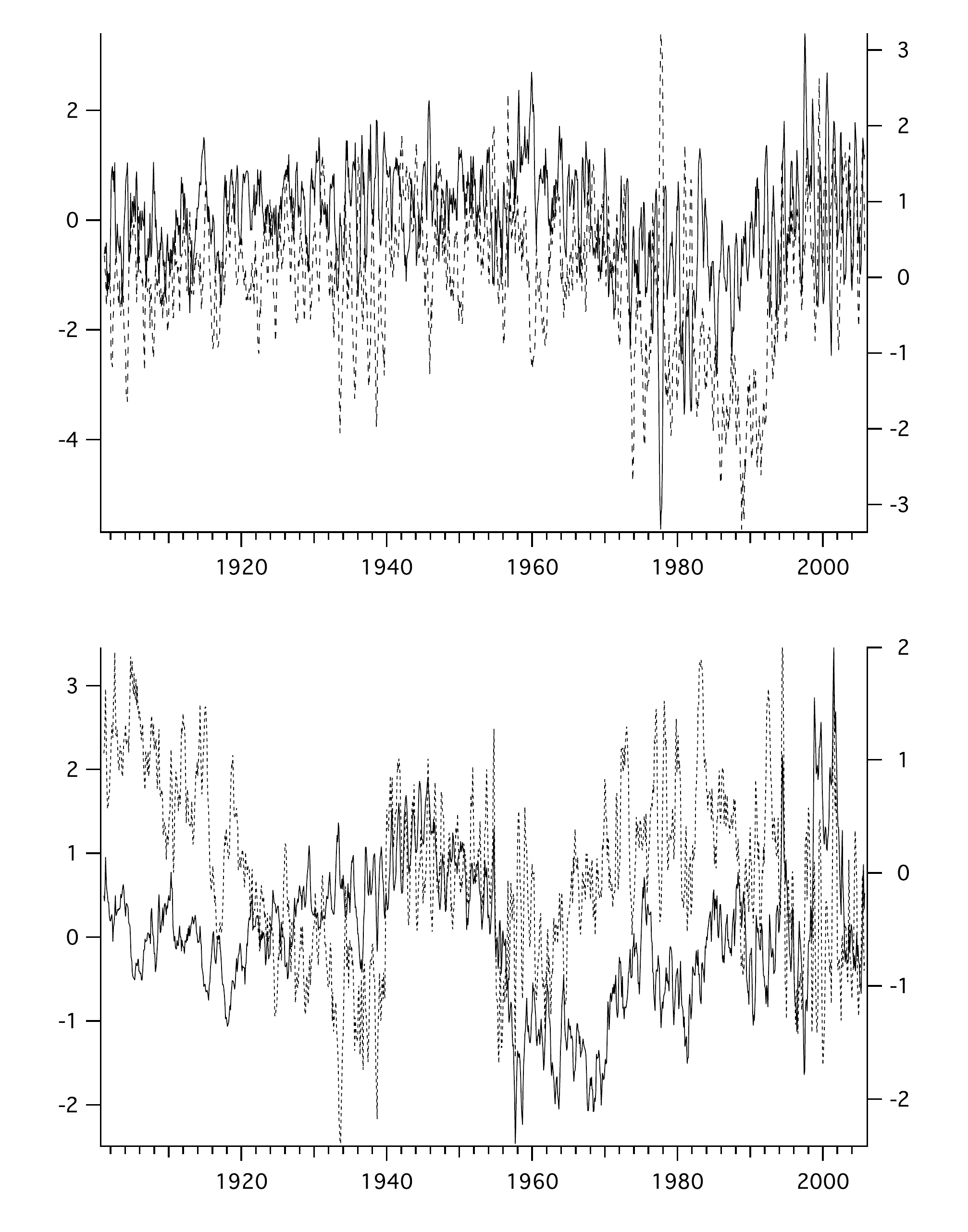}
 \caption{As in Figure~\ref{fig:common_trend1234} except for common trends 5-8.
}
 \label{fig:common_trend5678}
\end{figure}

\begin{figure}
\noindent  \includegraphics[width=6in]{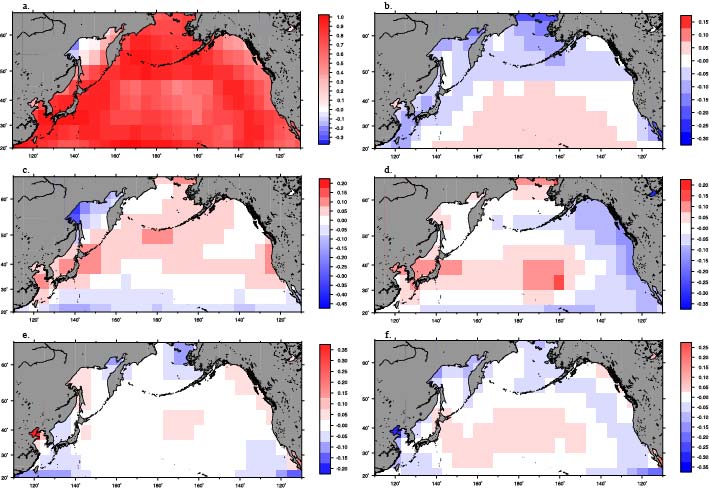}
 \caption{a). (Negative of) correlations between the univariate trends at each
location and common trend 1.  b).-f). Factor loadings by location for common
trends 2-6.}
 \label{fig:factor_loading}
\end{figure}

The (negative of the) first common trend ( see Figure~\ref{fig:common_trend1234}
for common trends 1-4, and Figure~\ref{fig:common_trend5678} for common trends
5-8) shows a sharp increase starting  in the late 1930's to 1940 period, which
reaches its maximum during 1956-1958, decreases till about 1984 and then
increases to the end of the series, with an acceleration occurring in roughly
1998.  The correlations between the univariate trends and the first common trend
(Figure~\ref{fig:factor_loading}a) shows that the first common trend
significantly correlates with the univariate trends in a band that is away from
the coast or North America, crosses the Pacific between around
50$^{\circ}$N-55$^{\circ}$N and 20$^{\circ}$N, abuts Japan but is weak in the
area of the West Wind Drift and the Kuroshio Extension.  The regions with
significant correlations have a factor loading with absolute value in the the
range of $0.10 - 0.07$ which translates into roughly a 0.8$^{\circ}$C - 
1.0$^{\circ}$C change over the 100-plus years analyzed.

The second common trend (Figure~\ref{fig:common_trend1234}) decreases slowly
till the early 1970's, when it declines, with the rate of decrease accelerating
around 1977-78 . 
The  significant factor loadings of the second common trend with the univariate
trends (Figure~\ref{fig:factor_loading}b ) only occur in a region between
20$^{\circ}$N-30$^{\circ}$N and 140$^{\circ}$W-160$^{\circ}$E and with opposite
sign in the Bering Sea. The "1976 regime shift"  (Hare and Mantua 2000;
Miller et al. 1994; Trenberth 1990 ) appears to be an acceleration
of a change that had already started in the North Pacific,  and was most
strongly expressed only in a limited area of the North Pacific, with warming in
the Bering Sea area and cooling in an area closer to the tropics after 1977-1978.
 This is consistent with results in \cite{schwing2007} who used
state-space models to examine a variety of physical and biological data along
the coast of North America, as well as some large-scale physical indices.

The first  two SST common trends display  significant long-term warming which
appears to be accelerating in more recent years, a trend reminiscent of global
warming signals. In order to see if this trend is unique to the North Pacific or
is more global in scope, we compare this result with a common trend analysis of
similar 5-degree SST summaries from the North Atlantic, with a common trend
analysis of 23 global pressure centers, based on the data in Minobe but with the
climatology restored, and with some indices of oceanographic conditions in the
tropics and the Arctic.   
 
\begin{figure}
\noindent  \includegraphics[width=6in]{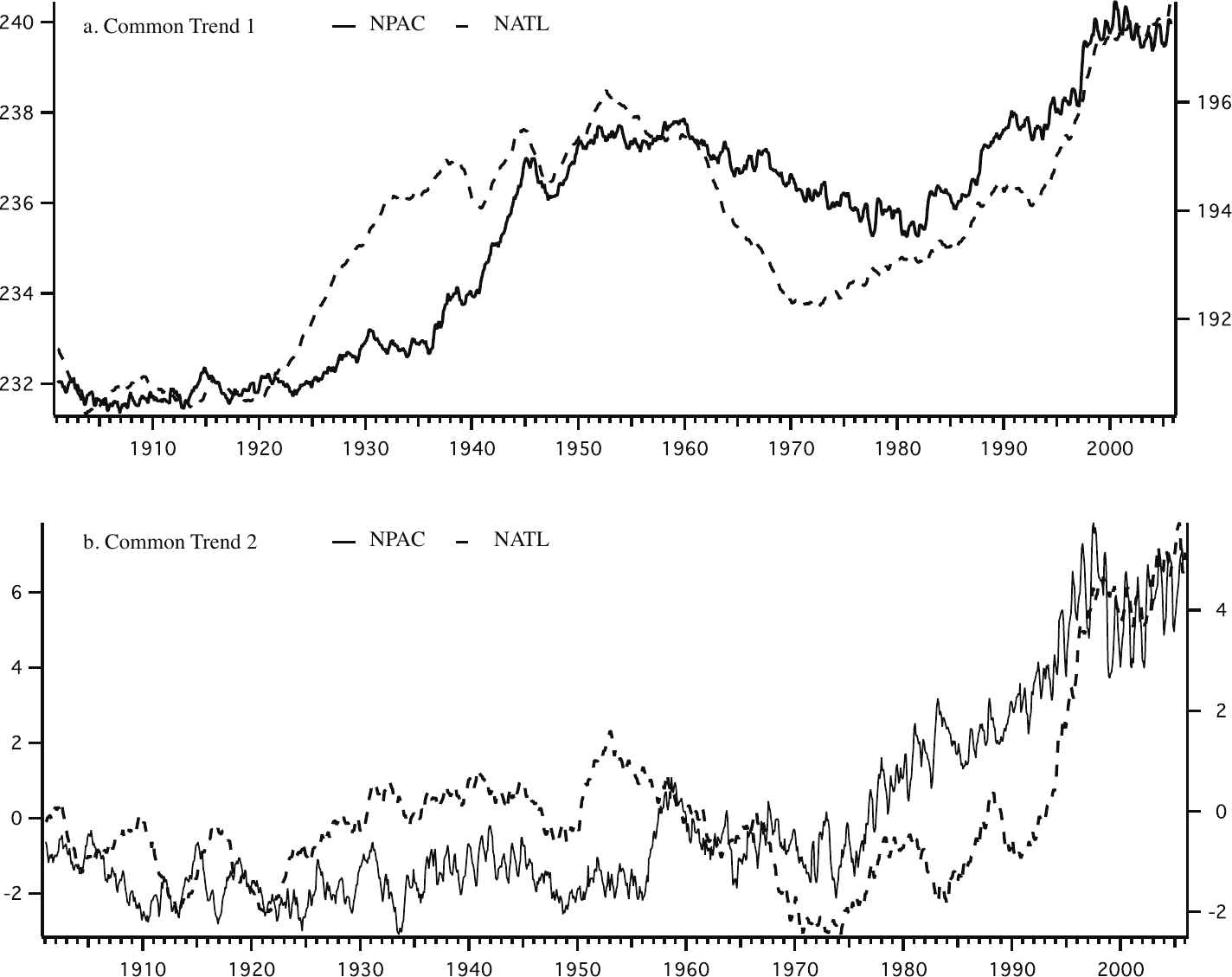}
 \caption{a). Comparison of the first common SST trends for the North Pacific
and the North Atlantic. b).Comparison of the second common SST trends for the
North Pacific and the North Atlantic }
 \label{fig:npac_natl_compare}
\end{figure}

The first two SST common trends  in the North Atlantic are similar to those from
the North Pacific  (Figure~\ref{fig:npac_natl_compare}). The factor loadings for
the North Atlantic first common trend are on the order of about .12, so that
during the same 100 year period the overall increase in temperature in the North
Atlantic is on the order 1.0$^{\circ}$C, as in the North Pacific.  Similar
results hold for the second common trends.

\begin{figure}
\noindent  \includegraphics[width=6in]{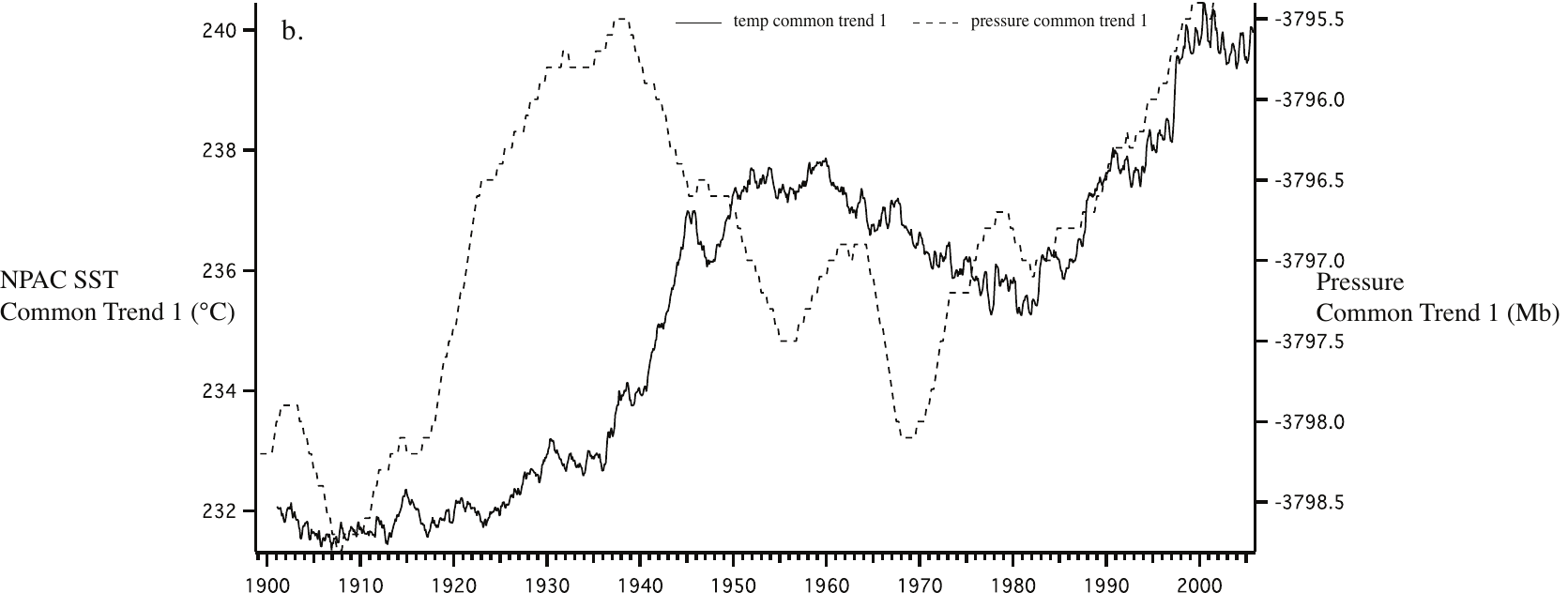}
 \caption{  Comparison of the negative of common trend 1  (solid line) with
common trend 1 from an analysis of 23 pressure centers  in  {\it Mendelssohn et
al.} (2007).}
 \label{fig:common1_pres1}
\end{figure}

The pressure common trend (Figure~\ref{fig:common1_pres1}b) has similar
change-points as does the North Pacific common trend, something first noted in
\cite{Schwing:2003zt}, with tight coupling after what appears to be a
major global shift around 1940.  Prior to that, while both series are
increasing, the rate of increase and other features of the series differ. 

\begin{figure}
\noindent  \includegraphics[width=6in]{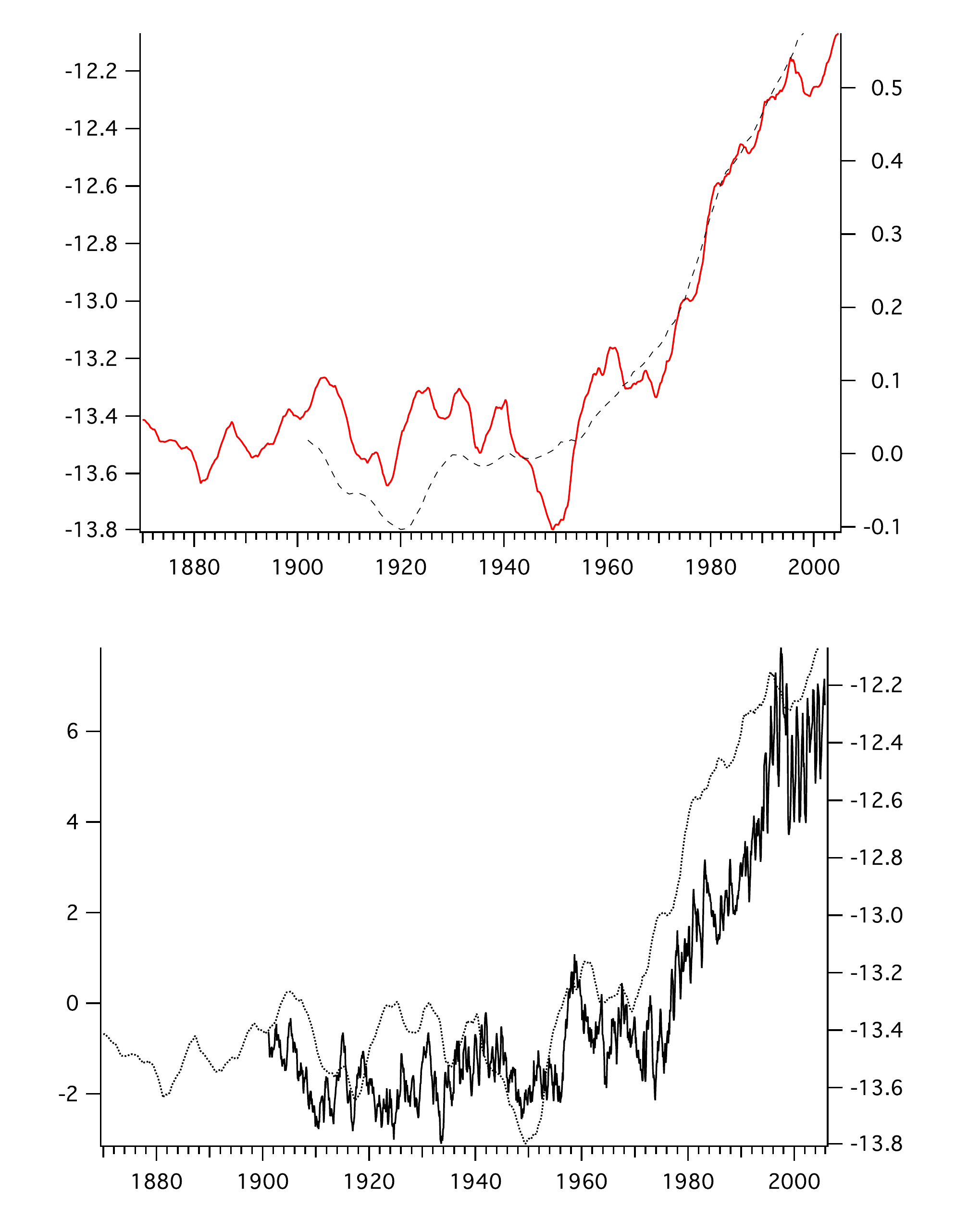}
 \caption{Comparison of the negative of common trend 2  (bold solid line) with
the trend for the Ni\~{n}o3 series  (dashed line) from {\it Mendelssohn et al.}
(2005) and the trend in North Pacific Sea Ice extent (light solid line).}
 \label{fig:common2_N}
\end{figure}

The (negative) of the second common trend is similar to the trend estimated by
\cite{ISI:000228479700003} for the yearly Ni\~{n}o3 index
(Figure~\ref{fig:common2_N}) as well to the trend estimated for the sea
ice extent in the Arctic, suggesting that this accelerating warming in more
recent decades  extends from the tropics to the Arctic. The Ni\~{n}o3 trend
appears to start its sharp increase earlier in time, suggesting the possibility
that the tropics both lead and influence the extra-tropical areas that load
significantly on the second common trend.  However, given that there is only one
such change-point in the series, it is impossible to test if this is a real or
chance relationship.

The third common trend (Figure~\ref{fig:common_trend1234}) shows a sharp
increase starting in 1940 till roughly 1956-1958, after which it decreases. The
factor loadings (Figure~\ref{fig:factor_loading}c ) for the third common trend
are most significantly positive along a portion of the west coast from central
California through Washington, along the Aleutians and along Japan, and most
negative in the Sea of Okhotsk and along a thin strip running roughly between
20$^{\circ}$N and 22$^{\circ}$N.

The univariate trends in the areas where the factor loadings are most positive
(not shown) have an increase starting around 1940 that is more pronounced, and a
decrease after 1956-1958 that is more rapid, than what would be predicted from
the first common trend alone.  These two change points align with sharp changes
in the pressure centers, as shown above and in \cite{Schwing:2003zt}, and
the areas with the most positive loadings are areas that would be affected by
major pressure systems - for example the NPH, Western North Pacific High, 
and the Aleutian Low.  The third common trend appears to be a superposition of a
more localized pressure related change in the ocean temperatures onto the global
signal represented by the first common trend.

\begin{figure}
\noindent  \includegraphics[width=6in]{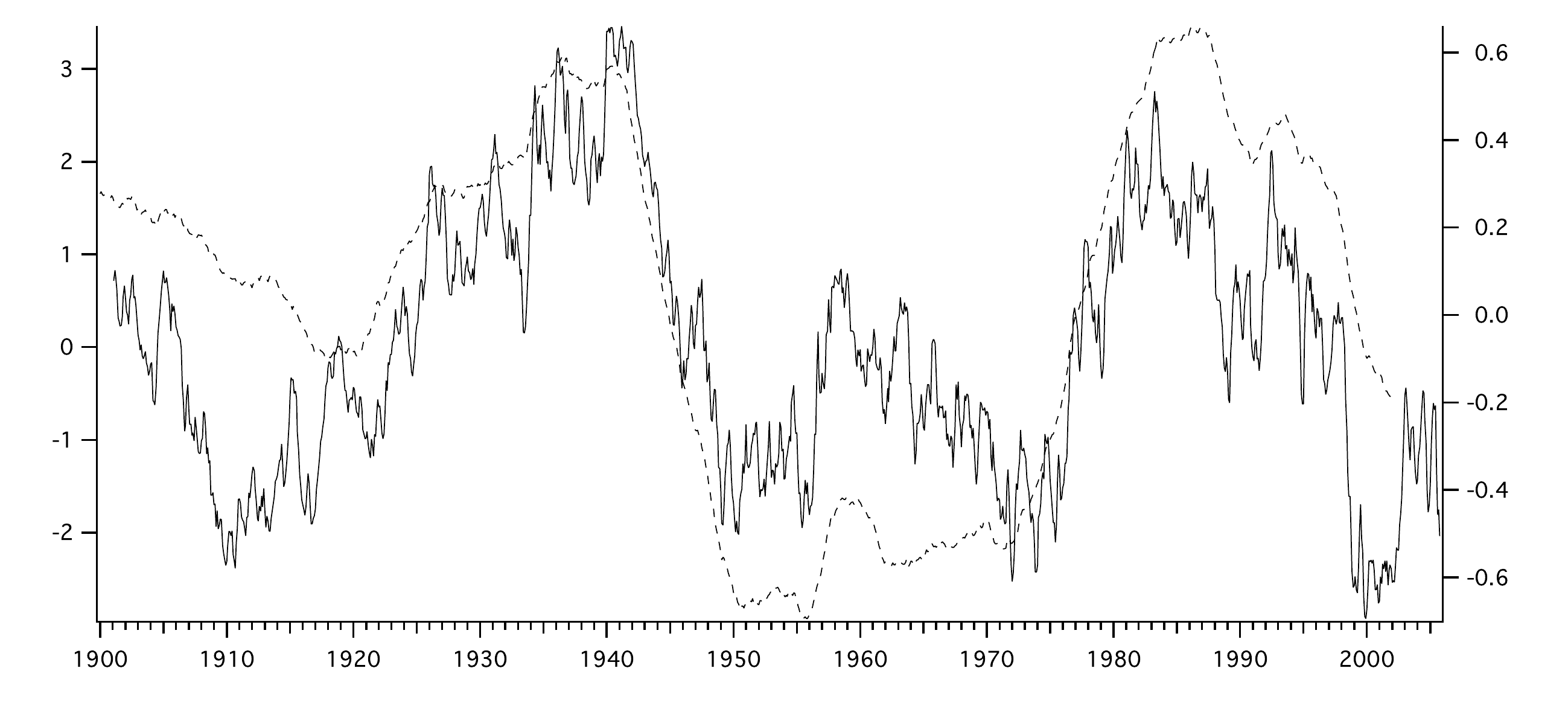}
 \caption{Comparison of the negative of common trend 4  (solid line) with the
trend for the PDO series from {\it Mendelssohn et al.} (2007).}
 \label{fig:common4_pdo}
\end{figure}

The fourth common trend (Figure~\ref{fig:common_trend1234}) loads most strongly
off the west coast of North America, the west wind drift area, and off Japan,
the first two being regions that should be affected by the NPH. 
The fourth common trend is almost identical to the trend estimated by
state-space analysis of the original PDO (Figure~\ref{fig:common4_pdo}). This
means that the long-term dynamics of the original PDO are capturing a more
localized behavior that is not the  dominant long-term dynamic in the North
Pacific.

\begin{figure}
 \noindent  \includegraphics[width=6in]{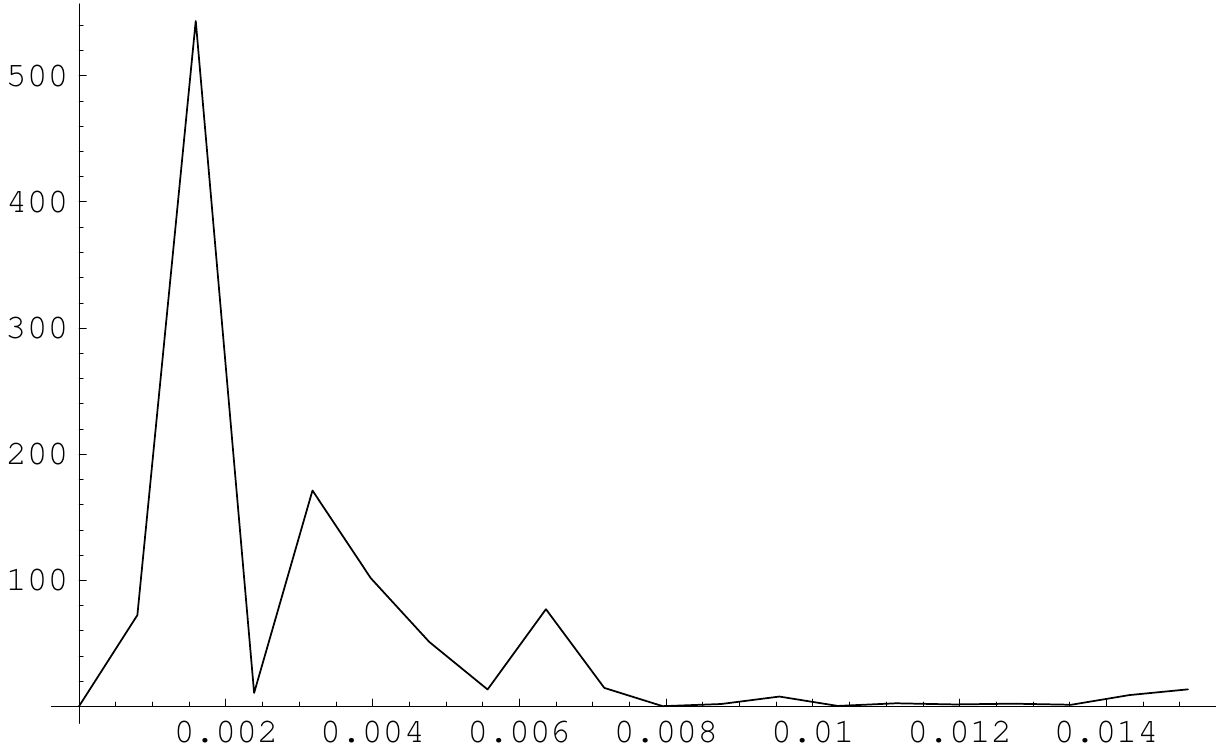}
\caption{Power spectrum of the fourth common trend.}
 \label{fig:common4_power}
\end{figure}

The spectral behavior of the fourth common trend
(Figure~\ref{fig:common4_power}) has peaks at 52, 26 and 13 years.  These
periods are the harmonics of the so-called "Seuss Wiggles"  \citep{Suess1965, Thomson24041990} noticed in the analysis of ice cores, and which appear to
reflect solar cycles.  This suggests the very plausible link between solar
cycles, changes in pressure systems in the atmosphere, and some of the
variability seen in the ocean. 

Common trends 5 through 8 (Figure~\ref{fig:common_trend5678}) are shown for
completeness.  There are quasi-cyclic series whose factor loadings
(Figure~\ref{fig:factor_loading}e-h) are very localized and do not represent
broad-scale changes in the North Pacific.  However, as we will show below, for
those regions where the factor loadings are of relatively greater absolute
value, these higher order common trends still play an important role in
reconstructing the dynamics in those regions.

\subsection{Is Variance Enough?}
\label{sec:R.variance}

When studying  decadal variability over a large spatial area, the main interest
is often change-points or others signs of shifts in the dynamics away from what
had been quasi-stationary behavior.  An EOF analysis is based on the most
variation in the series, while a common trends analysis attempts to find common
movements in the series, which would  seem most likely to correctly identify
change-points and shifts.

We have shown elsewhere that when the univariate trends are reconstructed from a
too low-dimension basis  (say only the first or first two modes) that many
important features of the original series are missed \citep{Mendelssohn:2002jl, Mendelssohn:2003um}, and the same is true for the
North Pacific SST.   The weightings of the common trends make it clear that
different parts of the ocean are behaving differently, so that it is impossible
to capture all of the dynamics in a very low-dimensional space. This is similar
to the famous regression example of  \cite{anscombe:1973} who shows four
regressions that have identical statistics, including $r^{2}$, but only one of
the regressions is an adequate fit to the data. In  our sub-space identification
procedure the goal is to finds the smallest underlying basis that still
adequately reproduces the important features of the original series.

\begin{figure}
\noindent  \includegraphics[width=6in]{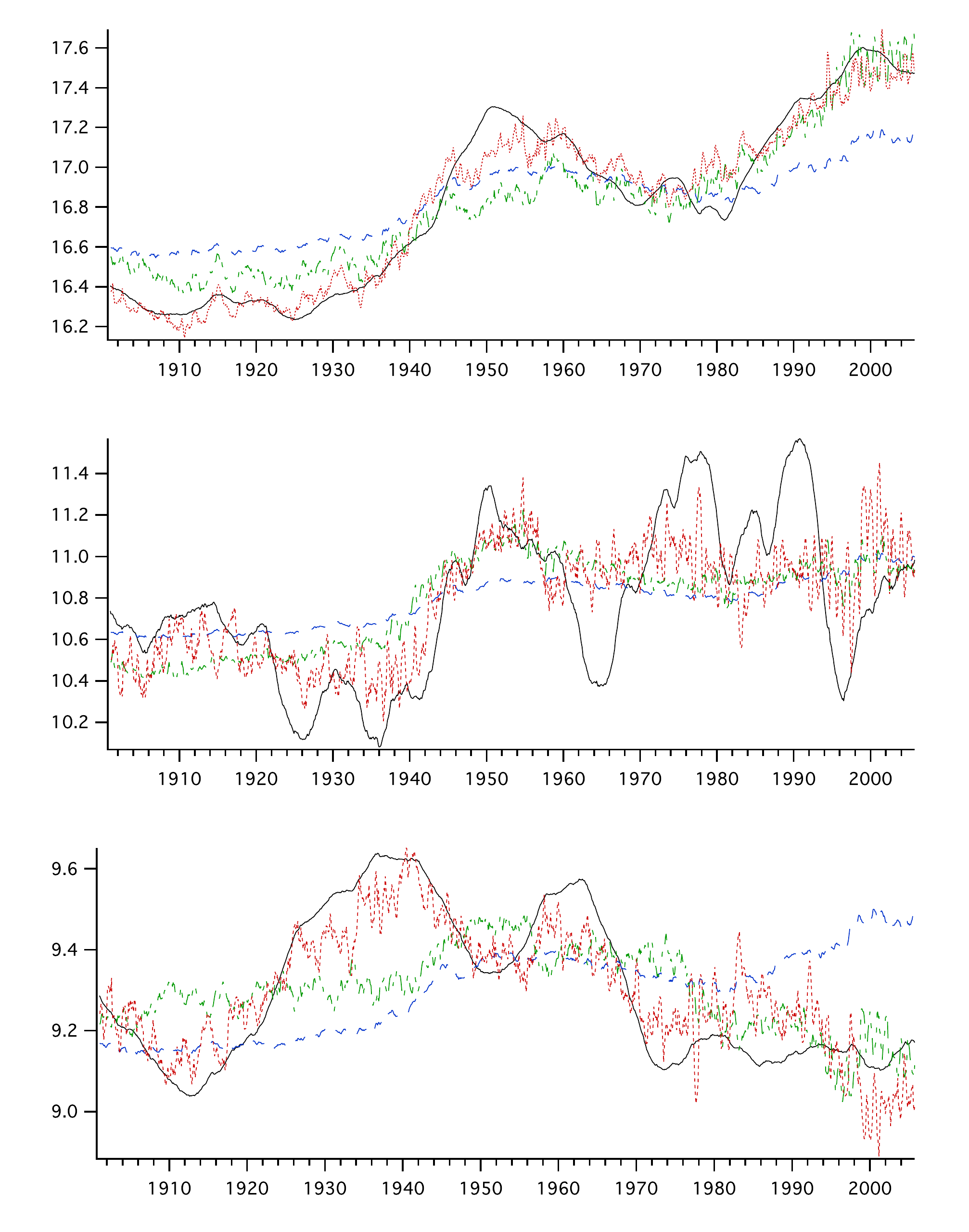}
 \caption{Reconstruction of a). the univariate trend at 30N, 120E (black,
$r$=-0.975) from common trend 1 (blue, common trends 1-2 (green) and common
trends 1-3 (red); b). the univariate trend at 40N, 130E f(black,$r$=-0.496) from
common trend 1(blue), common trends 1-3 (green), and common trends 1-5 (red);
c). the univariate trend at 55N, 135W (black,$r$=0.036) from common trend
1(blue), common trends 1-3 (green), and common trends 1-5 (red).  The $r$ values
are the correlation between the univariate trend and common trend 1.}
 \label{fig:area_recons}
\end{figure}

The univariate trend at 30$^{\circ}$N, 120$^{\circ}$E has a correlation of
$-0.975$ ($r^{2}= 0.95$) with the first common trend.  Despite this, if we
reconstruct the univariate trend from the first common trend
(Figure~\ref{fig:area_recons}a, blue line) it severely under-estimates the sharp
increase around 1950 and the large rise starting in 1992.  Adding the second
common trend (green line) captures these latter features, and it is not until
the first three common trends are used (red line) that the original trend is
reproduced.  Thus, despite the extremely high correlation between the univariate
trend in this region and the first common trend, reproducing the univariate
trend using only the first common trend would not give an accurate impression of
two of the most important features of the original trend.

In many analyses, a large number of the series have a medium level of
correlation with the first mode.  The univariate trend at 40$^{\circ}$N,
130$^{\circ}$E has a correlation of $-0.496$ ($r^{2}= 0.245$) with the first
common trend.  The reconstruction of the univariate trend using the first common
trend (Figure~\ref{fig:area_recons}b, blue line) fails to reproduce any of the
significant features of the univariate trend. 
Reconstructing using the first three common trends
(Figure~\ref{fig:area_recons}b, green line) captures the mean shift in the late
1930's-1940, but not the variation around the two levels, and even using the
first five common trends (Figure~\ref{fig:area_recons}b, red line) only
partially captures the variation post-1950.

As a final example, at 55$^{\circ}$N, 135$^{\circ}$W the univariate series has 
a correlation of $0.036$ ($r^{2}= 0.001$) with the first common trend.  This
area is up in the eastern part of the Gulf of Alaska, and because of the shape
of the coast, is one of the few areas in the region that is truly coastal.  
\cite{Mendelssohn:2003um} and \cite{Palacios:2004km} show that along the
west coast of North America there is a sharp onshore-offshore gradient in almost
every major feature including mixed layer depth, stratification, integrated heat
content etc. at scales much smaller than the 5$^{\circ}$ boxes used here, and 
it is questionable that any analysis using 5$^{\circ}$ boxes will accurately
capture coastal dynamics. This may explain why the sets of results at differing
resolutions differ, and underline the fact that coastal processes may not be
well represented in more broadscale analyses. This region is one of the few
regions in the North Pacific with a negative correlation  with the first common
trend, and in fact the first common trend does a poor job of reconstructing the
univariate trend  (Figure~\ref{fig:area_recons}c, blue line). 

Reproducing the univariate trend from the first three common trends
(Figure~\ref{fig:area_recons}c, green line) somewhat captures the decline
starting in the late 1950's or 1960, but it is not till the first five common
trends are used (Figure~\ref{fig:area_recons}c, red line) that the dynamics of 
the univariate trend is well reproduced. 

The reconstructions of the univariate trends in these three regions provide clear
examples that variance can be misleading in determining if a mode reproduces
change-points and other features of time series, and that even if a mode
explains a reasonable amount of the variance, there may be large areas in the
ocean that are not explained well by it.   In the case of the original PDO, its
trend represents two distinct regions in the North Pacific that have strong
interactions with the variation in the NPH,  but does not reflect the trends in
the rest of the North Pacific.

\subsection{Common Stochastic Cycles}
\label{sec:R.cycles}

The analysis of common stochastic cycles proceeds analogously to the analysis 
of common trends 
except that the stochastic cycles from the univariate state-space decompositions
are used rather than the trends. Stochastic cycles, by definition, are stationary
with a
constant mean and covariance structure, so the dynamics of the cycles are not
changing with time.  These types of dynamics are somewhat separate from the main
focus in this paper, 
the long-term, non-stationary behavior of North Pacific SST which will reflect
 "climate change" and other change-points in the data. However we examine the
common stochastic cycles in order to further
understand the role of "red noise" in the dynamics of NPAC SST,  
why \cite{ISI:000171261600018} and \cite{ISI:000188458400001} find that 
the PDO is essentially red noise, and to show other well know phenomena in the
Pacific region appear to be related to the stationary, rather than the
non-stationary dynamics of SST.. 

\begin{figure}
 \noindent  \includegraphics[width=6in]{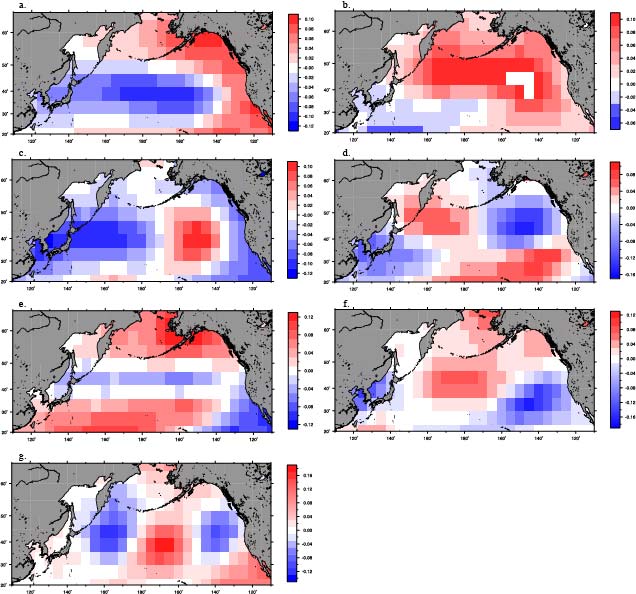}
\caption{a).-f). Factor loadings by location for common cycles 1-6.}
 \label{fig:factor_cycles}
\end{figure}

\begin{figure}
\noindent  \includegraphics[width=6in]{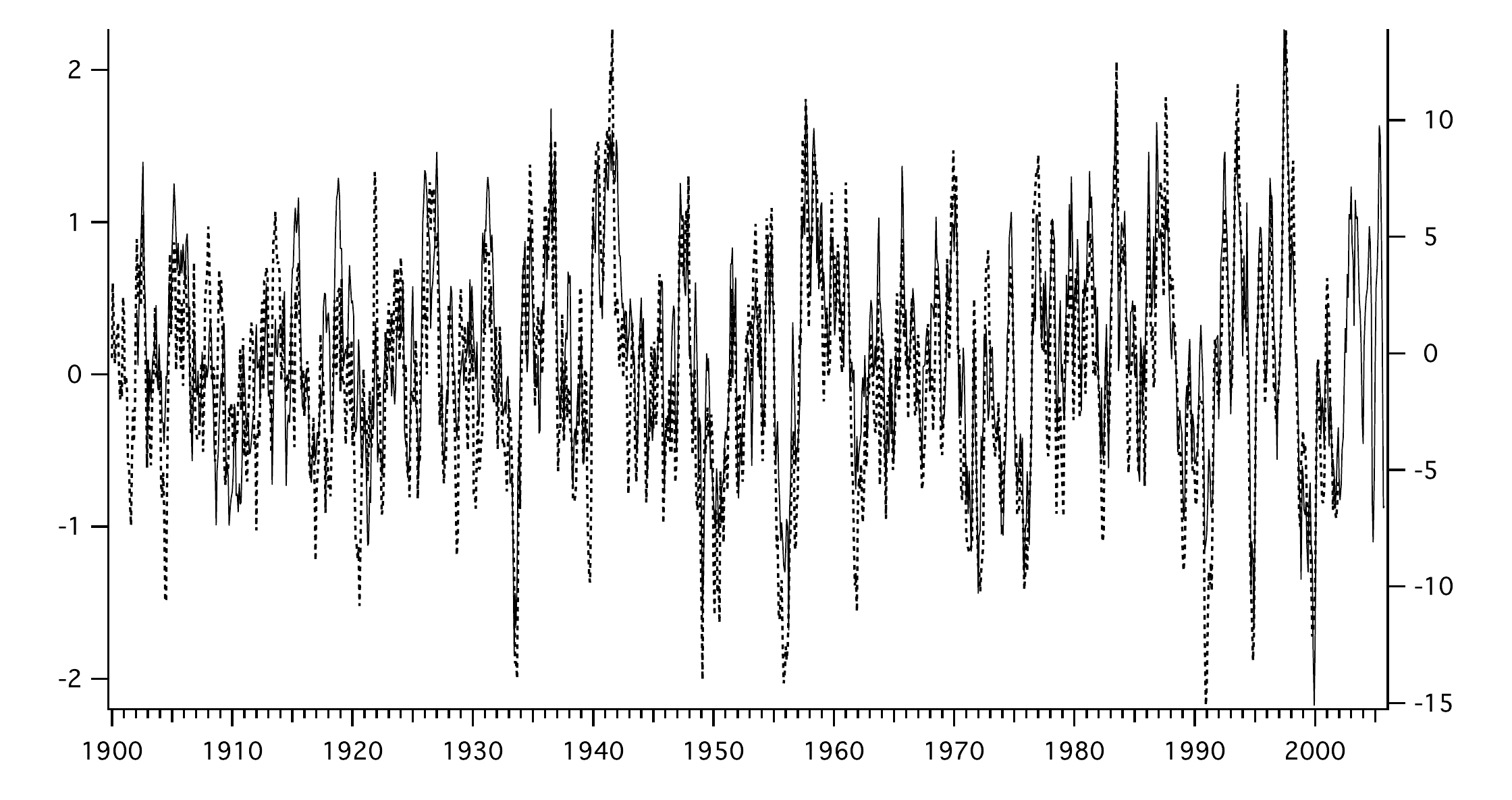}
 \caption{Comparison of the first common cycle (solid line) and the stochastic
cycle estimated from a state-decomposition of the PDO}
 \label{fig:cyc1_pdo_cyc}
\end{figure}

The spatial pattern of the factor loadings for the first two common stochastic
cycles (Figure~\ref{fig:factor_cycles}a,b) are very similar to the spatial
patterns for the original PDO and the so-called "Victoria mode" (Bond et
al. 2003) the second principal component of the SST anomalies. In fact, the
first common stochastic cycle is almost identical to the stochastic cycle
calculated from a state-space decomposition of the original PDO
(Figure~\ref{fig:cyc1_pdo_cyc}). Thus, while there is a "trend" term buried in
the original PDO, most of the analyses, including the spatial  and temporal
behavior, have focused on the stationary aspects of the PDO.  Moreover, as much
of the region that loads strongly on our first common cycle also loads strongly
on our fourth common trend, and both include the regions that should be affected
by the North Pacific High, we can see that it is this spatial overlap that
produced our original state-space decomposition of the PDO. When an EOF is
calculated from the anomalies, the stationary component contains more variance
than does the trend term since it is at a higher frequency, and since the two
regions have a high degree of overlap, the trend from this same region is
indirectly 
included in the EOF. 

The common cycles can be analyzed to ascertain the implied frequency of the
term. The first common cycle has a period of 3.62 years, the second common cycle
has a period of 11.4 years and the fourth common cycle has a period of 26
months. The periodicity of the first common cycle is consistent with the periods
of the stochastic cycles for Darwin and Tahiti sea-level pressures estimated in
\cite{ISI:000228479700003}.  However, for a variety of reasons, this
similarity in the period of the stochastic cycle is believed due to similarity
in the dynamics of the dominant pressure centers throughout the tropical and
extra-tropical north Pacific, rather than the tropics driving the extra-tropical
temperatures.. The period of the second common cycle is consistent with the
periodicity of solar cycles, in particular sun spots, and the periodicity of the
fourth common cycle is the quasi-biennial cycle. Thus the dominant stationary
cyclic behavior of North Pacific SST is the occurring at the same periods as
that of the atmosphere.

The spatial pattern of the factor loadings for common stochastic cycles 3-7
(Figure~\ref{fig:factor_cycles}c-g) are similar to those expected from dynamics driven by spherical harmonics.  The factor loadings show large spatial regions in a single
direction that vary out of phase with each other as the cycle progresses through
time, giving the appearance of warmer or cooler waters moving through time in
that direction. For the third common stochastic cycle there are  two regions out of phase in the
east-west direction while the fourth common stochastic cycle also 
has two regions out of phase but in the southwest-northeast and northwest-southeast
directions.

Similarly, for the sixth common stochastic cycle there are three regions with
alternating phases 
in the east-west direction and for the seventh common stochastic cycle there are
three regions alternating out of phase from the northwest- central-northeast
regions of the Pacific. The implications of this spherical harmonic type
dynamics in the North Pacific is a topic of further research.

\section{Discussion and Summary}
\label{sec:DS}

We have presented a new analysis of North Pacific SST that paints a very
different picture of the long-term dynamics in this region.  The "standard"
picture is based on the PDO, which we have shown that as originally calculated
captures mostly the stationary, rather than non-stationary behavior of only a
part of the North Pacific. The so-called "Victoria mode" is also a stationary
series that likewise reflects only  part of the North Pacific. These two series
are reproduced as our 
first two common stochastic cycles. Neither of these series represent the
long-term changes in the mean or variance that we might mean when we discuss
"regime changes", because as predominantly stationary series they have a
constant mean and constant covariance structure.  The changes attributed to the
PDO index, as presently calculated,  reflect an over-interpretation of the
dynamics of autoregressive-type series, which, like lotteries, can produce
regime-like behavior (long runs of numbers) without any real change in the
underlying mechanism.

The long-term non-stationary dynamics estimated in our state-space analysis
defines common trends that do contain important change-points that reflect
changes seen in other parts of the global environment.  Our first common trend
behaves like a "global warming" 
signal, and we show that SST in the North Atlantic have a similar first common
trend, suggesting the global nature of this trend. The first common trend 
also displays many similarities to the first common trend from an analysis of
pressure centers, and suggest that rather than 1976, both the North Pacific
temperatures and the pressure systems showed changed points around 1910, around
1920, in the late 1930's-1940, in the late 1950's to early 1960's , and in the
period from about 1982-1984.  These are the periods that appear to reflect major
reorganizations in the North Pacific system, but surprisingly are hardly
discussed in the literature, and for the most part agree with the analysis of
abrupt changes in \cite{Schwing:2003zt}.

Our second common trend represents the "1976 regime change"  as being an
acceleration of a change that had began previously  (the early 1970's)   and
which is an important event in the Bering Sea region and a near-equatorial band,
similar to results found in \cite{schwing2007}.  The second common
trend also  is similar to the estimated trend for the Ni\~{n}o3 series in
Mendelssohn et al. (2005), with a suggestion of a lagged relationship between
changes in the tropics and changes in these regions.

Hidden in the original PDO is a trend that is identical to the trend in the
North Pacific High, and is captured as our fourth common trend.  It is important
in two areas, including near the coast of North America, that are strongly
influenced by the NPH.  This result also clarifies  what has happened in the
calculations of the PDO.  As higher frequency dynamics usually contain higher
variance than lower frequency dynamics, the PDO has captured the area that has
the highest variance in the stochastic cycles, but this region mostly coincides
with the region whose trend is related to the trend NPH.  The fourth common
trend has a spectrum with significant peaks at 52.26 and 13 years, which
coincide with the so-called "Seuss wiggles",  suggesting a solar influence to
this component.

The factor loadings for common stochastic cycles 3-7 (again which are
stationary) suggest wavelike (spherical harmonic) phenomena in the stochastic cycles, with large,
adjacent areas of the North Pacific varying out of phase with each other as the
stochastic cycles proceed.  This is an important area for future research.

Finally, from a methodological standpoint, we have presented very clear evidence
that the state-space techniques that we have utilized in this and other papers
do produce results that differ from those produced by an EOF analysis,  as the
original PDO calculations are an EOF analysis of the same temperature series
that we have analyzed  (the EOF analysis uses the anomalies).  Not just
differing results, but results that have important differences in identifying
when the ocean has displayed significant shifts, and what might be the dynamics
causing these shifts. 
Moreover, the dynamics that we have estimated using state-space techniques from
a wide variety of time series, 
from ocean temperatures to global pressures to biological data, are consistent
in both temporal and spatial behaviors.

\bibliographystyle{crs}
\bibliography{pdoReanalysis}

\newpage

\end{document}